\documentclass[aps,prl,twocolumn,superscriptaddress,amsmath,floatfix]{revtex4} 

\usepackage{natbib}
\usepackage[T1]{fontenc}
\usepackage[colorlinks,hyperindex,colorlinks,citecolor=blue,linkcolor=blue,urlcolor=blue,]{hyperref}   
\usepackage{cleveref}
\usepackage{array}
\usepackage{tabularx}
\usepackage{graphicx}  
\graphicspath{{./}}
\usepackage{subfigure}
\usepackage{epstopdf}
\usepackage{dcolumn}        
\usepackage{amssymb} 
\usepackage{amsmath}
\usepackage{booktabs}
\usepackage{multirow}
\usepackage{color}
\usepackage{upgreek}
\usepackage{textcomp}
\usepackage[marginal]{footmisc}

\usepackage{siunitx}

\makeindex

\begin{document}

\title{Ultralow thermal conductivity in two-dimensional MoO$_3$}

\author{Zhen Tong}
\affiliation{Shenzhen JL Computational Science and Applied Research Institute, Shenzhen 518110, China.}

\author{Traian Dumitric\u{a}}
\email{dtraian@umn.edu}
\affiliation{Department of Mechanical Engineering, University of Minnesota, Minnesota 55455, United States}

\author{Thomas Frauenheim}
\email{thomas.frauenheim@bccms.uni-bremen.de}
\affiliation{Shenzhen JL Computational Science and Applied Research Institute, Shenzhen 518110, China.}
\affiliation{Beijing Computational Science Research Center, Beijing 100193, China}
\affiliation{Bremen Center for Computational Materials Science, University of Bremen, Bremen 2835, Germany}

\date{\today}
\begin{abstract}
Monolayer molybdenum trioxide (MoO$_3$) is an emerging two-dimensional (2D) material with high electrical conductivity. Using first-principles calculations and a Boltzmann transport theoretical framework, we predict record low room-temperature phonon thermal conductivity ($\kappa_p$) of 1.57 W/mK and 1.26 W/mK along the principal in-plane directions of MoO$_3$ monolayer. The behavior is attributed to the combination of soft flexural and  in-plane acoustic modes, which are coupled through the finite layer thickness, and to the strong bonding anharmonicity, which gives rise to significant 3- and 4-phonon scattering events. These insights suggest new indicators for guiding the search of 2D materials with low $\kappa_p$. Our result  motivates experimental  $\kappa_p$ measurements in MoO$_3$, and its applications as a thermoelectric and thermally protective material. 

\end{abstract}

\maketitle


Two dimensional (2D) materials beyond graphene \cite{novoselov_electric_2004} are receiving widespread attention as their unique physical properties \cite{gu_colloquium_2018} can enable next-generation technologies \cite{kautzky_materials_2018}. In terms of their lattice thermal conductivity, $\kappa_p$, 2D materials can display \cite{gu_colloquium_2018}  both high $\kappa_{p}$ materials, such as graphene \cite{cai_thermal_2010}, BN \cite{cai_high_2019}, MoS$_2$ \cite{li_thermal_2013}, and WS$_2$ \cite{gu_phonon_2014}, and low $\kappa_p$, such as SnSe \cite{qin_diverse_2016,villanova_anomalous_2021}, SnS \cite{qin_diverse_2016}, phosphorene \cite{jain_strongly_2015}, silicene \cite{xie_thermal_2014}, and tellurene \cite{gao_unusually_2018}.  Materials from the latter category are of interest for Peltier refrigerators and Seebeck power generators, where low $\kappa_p$ can enable a high figure of merit $ZT=S^2\sigma T/(\kappa_{p}+\kappa_{e})$.  Here $S$ is the Seebeck coefficient, $\sigma$  the electrical conductivity,  $\kappa_{e}$ the electronic contribution to thermal conductivity, and $T$ the absolute temperature. 
In addition to low $\kappa_p$, good stability  in oxidative air-bearing environments is often a prerequisite for the various applications. For example, although phosphorene has a low $\kappa_p$ and large $ZT$, it is susceptible to oxidization due to the lone-pair electrons in each phosphorus atom \cite{C4NR05384B}.  Low $\kappa_p$ nanostructures are also needed for engineering protective barriers in nm-spacings subjected to the extreme thermal constraints \cite{kautzky_materials_2018} often encountered in nano and micro-scale devices. 

\begin{figure}[htp]
\subfigure{\includegraphics[width=7 cm]{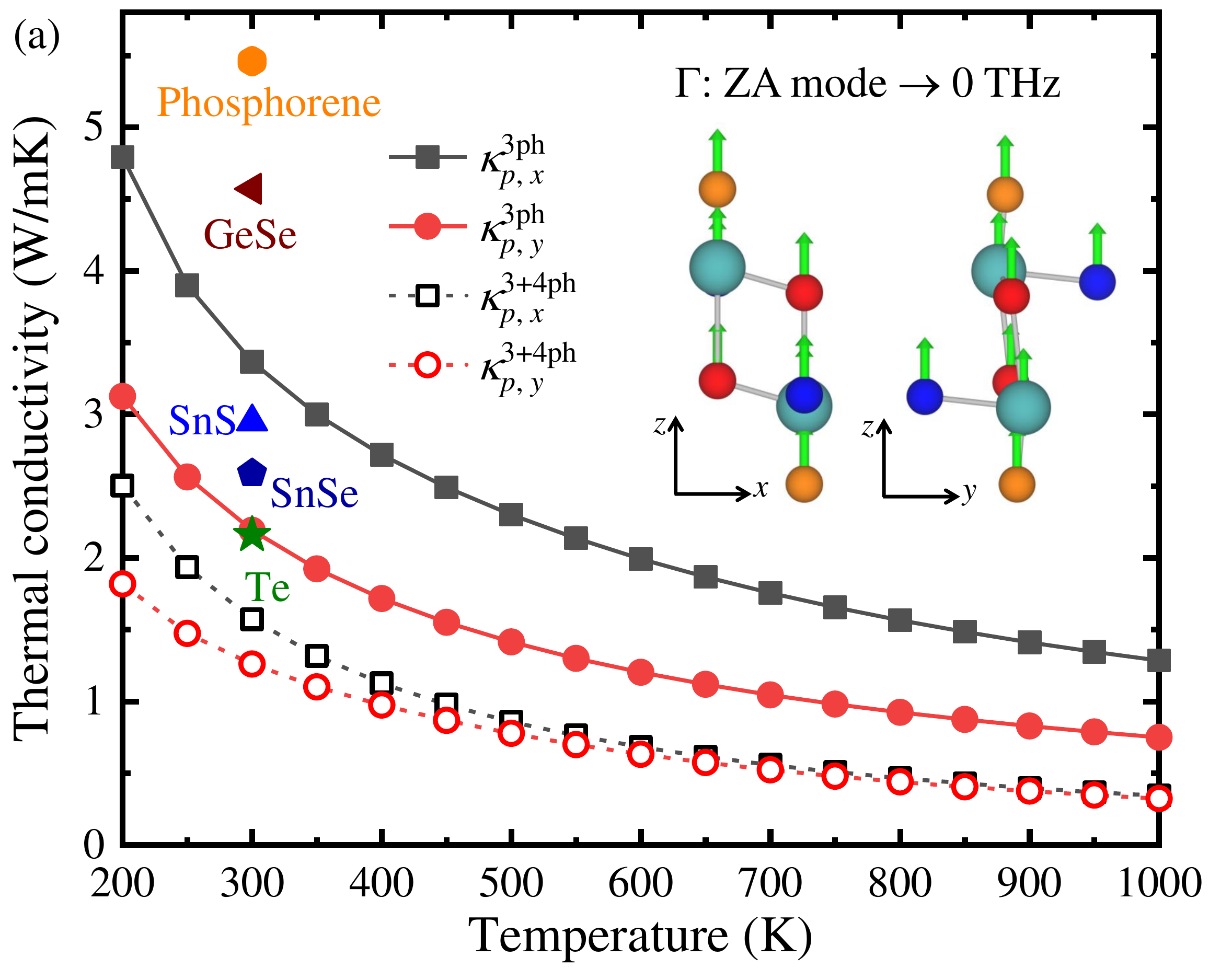}}
\subfigure{\includegraphics[width=7 cm]{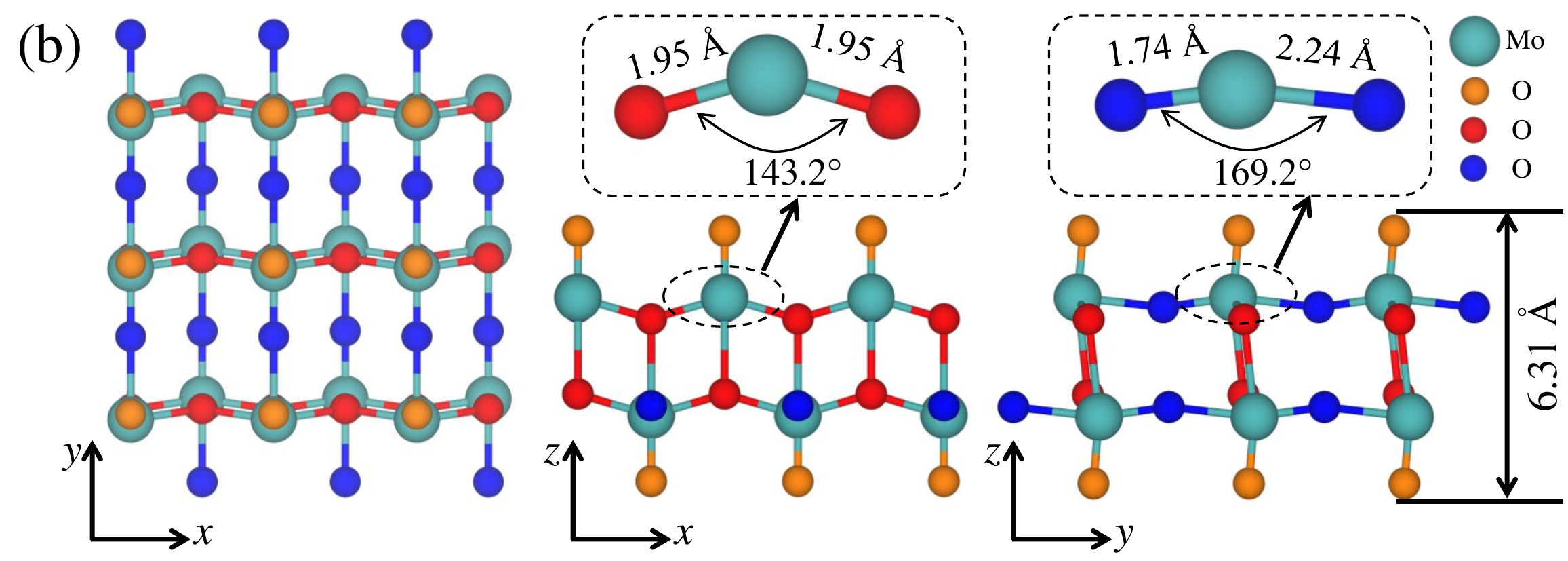}}
	\caption{\label{fig:1}
	(a)  $\kappa_p$ of monolayer MoO$_3$ vs. $T$.  $\kappa^{\text{3ph}}_{p}$ and $\kappa^{\text{3+4ph}}_{p}$ along  $x$ and $y$ directions (as indicated in the subscript) are also plotted. Superscript  notation  denotes  inclusion of 3-phonon only, and both 3- and 4-phonon scattering events.  Literature data (theoretical) for 2D SnSe, SnS, GeSe \cite{qin_diverse_2016}, phosphorene \cite{jain_strongly_2015}, and tellurene \cite{gao_unusually_2018}. {Inset shows the  ZA vibrational pattern near $\Gamma$ point.} (b) Monolayer MoO$_3$ in top and side views.}
\end{figure}

In this Letter we investigate  $\kappa_p$ of 2D MoO$_3$, the compound of Mo at its highest oxidation state, MoO$_3$.  Bulk MoO$_3$ is currently utilized in a variety of applications ranging from infrared detection \cite{foteinopoulou_phonon-polaritonics_2019}, catalysis \cite{de_castro_molybdenum_2017},  to solar cells \cite{girotto_solution-processed_2011} and field electronics \cite{alsaif_high-performance_2016}. On the fundamental side,  extensive works have been carried out to explore its unique properties. For example, Ma $et$ $al$. \cite{ma_-plane_2018} reported anisotropic in-plane phonon polaritons, while Chen $et$ $al$. \cite{chen_configurable_2020} and Hu $et$ $al$. \cite{hu_topological_2020} reported phonon and topological polaritons in twisted MoO$_3$.  Negishi $et$ $al$. \cite{negishi_anisotropic_2004} measured the thermal expansion coefficients of crystalline MoO$_3$,  while Kaiser $et$ $al$. \cite{kaiser_molybdenum_2020} recently measured a low $\kappa_p$ of 3.28 W/mK in powder MoO$_3$.   2D MoO$_3$ was succesfsully exfoliated from the crystalline form comprising orthorhombic monolayers bonded by  van der Waals (vdW) forces \cite{balendhran_enhanced_2013,balendhran_field_2013,alsaif_high-performance_2016,kalantar-zadeh_synthesis_2010,de_castro_molybdenum_2017}. Although 2D MoO$_3$ has been predicted to display high electronic mobility \cite{balendhran_enhanced_2013,zhang_high-mobility_2017}, its  $\kappa_p$ remains unexplored. Relying on vdW corrected density functional theory (DFT) calculations, here we report record low $\kappa_p$ in the MoO$_3$ monolayer.


We begin by describing our theoretical framework. 
Combining Boltzmann transport equation with Fourier's law of heat conduction \cite{mahan_many-particle_2000}, the elements of the phonon thermal conductivity tensor   write as a summation over all phonon modes  $\lambda  = \left( {{\rm{\bf{q}},\nu}} \right)$  of polarization $\nu$, wave vector $\rm{\bf{q}}$ and frequency $\omega_{\lambda}$ \cite{tong_decompose_2018,tong_comprehensive_2019,li_anomalous_2020}
\begin{equation}\label{eq1}
\kappa _{p,\alpha \beta } =\frac{1}{N_{\bf{q}}} \sum\limits_\lambda  {{c_{\lambda }}} v_{\lambda,\alpha} v_{\lambda,\beta} {\tau _\lambda }.
\end{equation}  
Here, $\alpha$ and $\beta$ are indexing the Cartesian directions,  $N_{\bf{q}}$ is the total number of $\bf{q}$-points sampled in the first Brillouin zone, while  $c_{\lambda }$, $v_{\lambda}$ and $\tau_{\lambda}$ denote the volumetric heat capacity, phonon group velocity, and phonon relaxation time, respectively. Note that $c_{\lambda }=({\hbar\omega_{\lambda}}/{V})({\partial{n_{\lambda}^{0}}}/{\partial{T}})$, where $n_{\lambda}^{0}$ is the Bose-Einstein distribution function and $V$ the volume of the primitive cell, and  ${v}_{\lambda,\alpha}={\partial{\omega_{\lambda}}}/{\partial{{q_{\alpha}}}}$. Phonons can be scattered on other phonons, electrons, impurities, or grain boundaries. With the Matthiessen's rule \cite{mahan_many-particle_2000}, the phonon scattering rate ($1/\tau_{\lambda}$) is a summation of the  phonon-phonon ($1/\tau_{\lambda}^{\text{ph-ph}}$), phonon-electron ($1/\tau_{\lambda}^{\text{ph-e}}$), phonon-impurity ($1/\tau_{\lambda}^{\text{ph-im}}$), and phonon-grain boundary ($1/\tau_{\lambda}^{\text{ph-gb}}$) scattering rates.
Here we predict the $\kappa_p$ of monolayer MoO$_3$  based on {\it ab initio} computed $1/\tau_{\lambda}^{\text{ph-ph}}$  with 3- and 4-phonon scattering rates
\begin{equation}
\label{eq2}
\frac{1}{\tau_{\lambda}^{\text{ph-ph}}}\approx\frac{1}{{\tau _\lambda ^{\text{3ph}}}} + \frac{1}{{\tau _\lambda ^{\text{4ph}}}},
\end{equation}
which are accounted for by summing the probabilities for all possible 3- and 4-phonon scattering events~\cite{mahan_many-particle_2000}.  

Figure \ref{fig:1}(a) presents our central result, the calculated ultralow $\kappa_{p}$ along the principal in-plane directions.  Remarkably, the computed values are below those of the  previously reported for 2D SnSe, SnS, GeSe \cite{qin_diverse_2016}, phosphorene \cite{jain_strongly_2015} and tellurene \cite{gao_unusually_2018}. Our comprehensive analysis  fully supports this result and reveals its origins.

\begin{figure}[htp]
\subfigure{\includegraphics[width=7 cm]{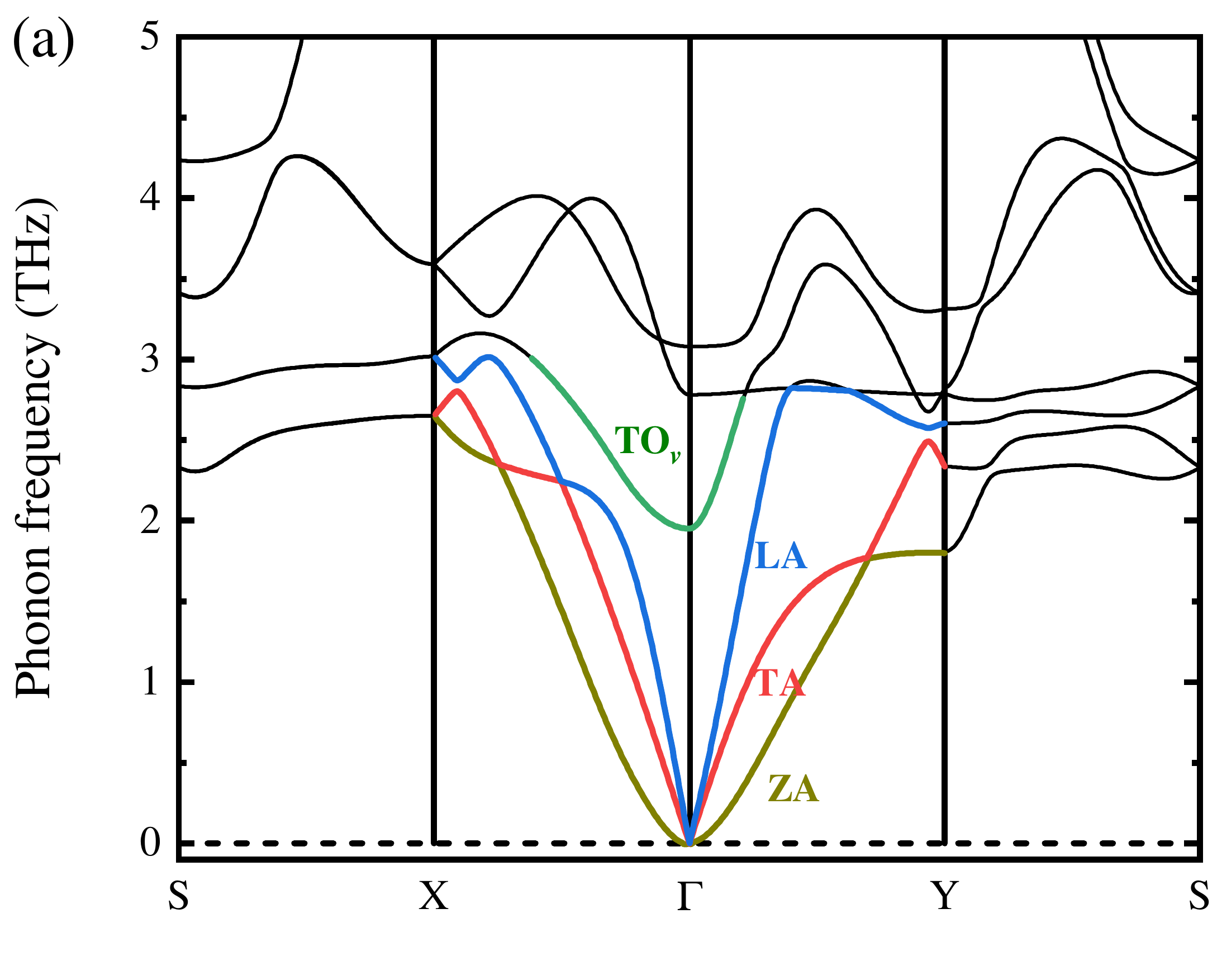}}
\subfigure{\includegraphics[width=7 cm]{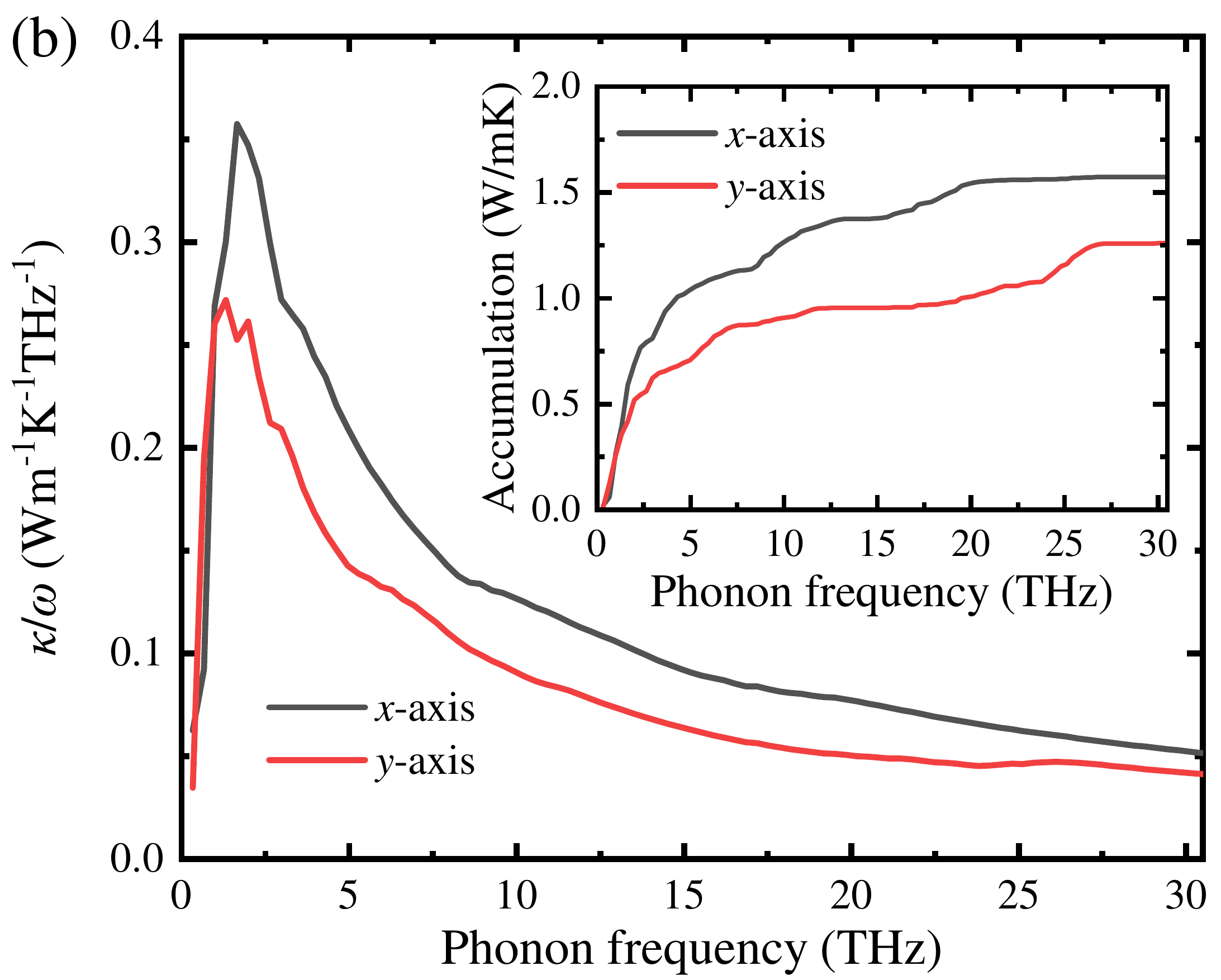}}
	\caption{\label{fig:2}
	(a) Phonon dispersion of 2D MoO$_3$ showing  acoustic  in-plane transverse (TA) and longitudinal (LA), and out-of-plane (ZA) modes, and an asymmetric valley-like transverse optical (TO$_v$) mode. (b) Frequency resolved $\kappa_p$ along $x$ and $y$ at $T=300$ K. Inset shows the  accumulation function.} 
\end{figure}

With 8 atoms per unit cell, monolayer MoO$_3$ has 3 acoustic and 21 optical phonon modes.  The phonon dispersion in 1$-$5 THz range, Fig. \ref{fig:2}(a), fully captures the acoustic modes. Near $\Gamma$, TA and LA are linear in $\bf{q}$, whereas ZA displays a quadratic trend. The latter is a typical feature for 2D materials and could be explained with elastic theory of thin plates \cite{gu_phonon_2014,gu_colloquium_2018}. Additionally, the ZA, TA, and LA  branches along $\Gamma$-X are steeper than along $\Gamma$-Y indicates a larger phonon group velocity along $x$. Specifically, at $\Gamma$ point we obtained $v_x$=3.19 km/s and $v_y$=3.54 km/s for TA, and $v_x$=3.80 km/s and $v_y$=1.07 km/s for LA.  A difference between $v_x$ and $v_y$ is consistent with the distinct $\kappa_{p}$ along $x$ and $y$, Fig. \ref{fig:1}(a).  Note also that  MoO$_3$ has structural anisotropy, as it can be seen in the Mo-O bond lengths and angles, Fig. \ref{fig:1}(b).  Along $x$-axis (O in red), the bonds measure 1.95 \AA{} and form $\theta_x$=143.2$^{\circ}$ angles.  Instead, along the $y$-axis (O in blue) the two distinct Mo-O bond types measure 2.24 \AA{} and 1.74 \AA{}, forming $\theta_y$=169.2$^{\circ}$ angles. Overall, the obtained small  $v_x$ and $v_y$ are strong indications for the intrinsically soft harmonic nature of the monolayer MoO$_3$. 


The phonon band structure of Fig. \ref{fig:2}(a) gives also the monolayer flexural rigidity ($D$) \cite{bending1_2011, liu_continuum_2016}.  Along $\Gamma$-X and $\Gamma$-Y directions, we obtained $D$=0.16 eV (along $x$) and $D$=0.11 eV (along $y$), respectively. These small $D$ are likely tied to energy changes through small bond angle ($\theta_x$ and $\theta_y$) variations, and are both much smaller than the 1.44 eV value of graphene \cite{bending1_2011}. We recall that the ultrahigh $\kappa_p$ of graphene originates in the decoupling of in-plane and out-of-plane phonon modes \cite{lindsay_phonon_2014,liu_continuum_2016} allowed by its one-atom-thick nature \cite{bending1_2011}. However, monolayer MoO$_3$ presents finite thickness,  {Fig. \ref{fig:1}(b)}, and should not exhibit such full decoupling.   



To focus the investigation, let us identify the important frequency range of heat-carrying phonons in 2D  MoO$_3$.  Figure \ref{fig:2}(b) shows the calculated frequency-resolved $\kappa_p$  while  the inset gives the $\kappa_p$ frequency accumulation (the cumulative frequency contribution to  $\kappa_p$). As in graphene, the low 0$-$5 THz frequency dominates $\kappa_p$.  

\begin{figure}[htp]
\centering     
\subfigure{\includegraphics[width=7 cm]{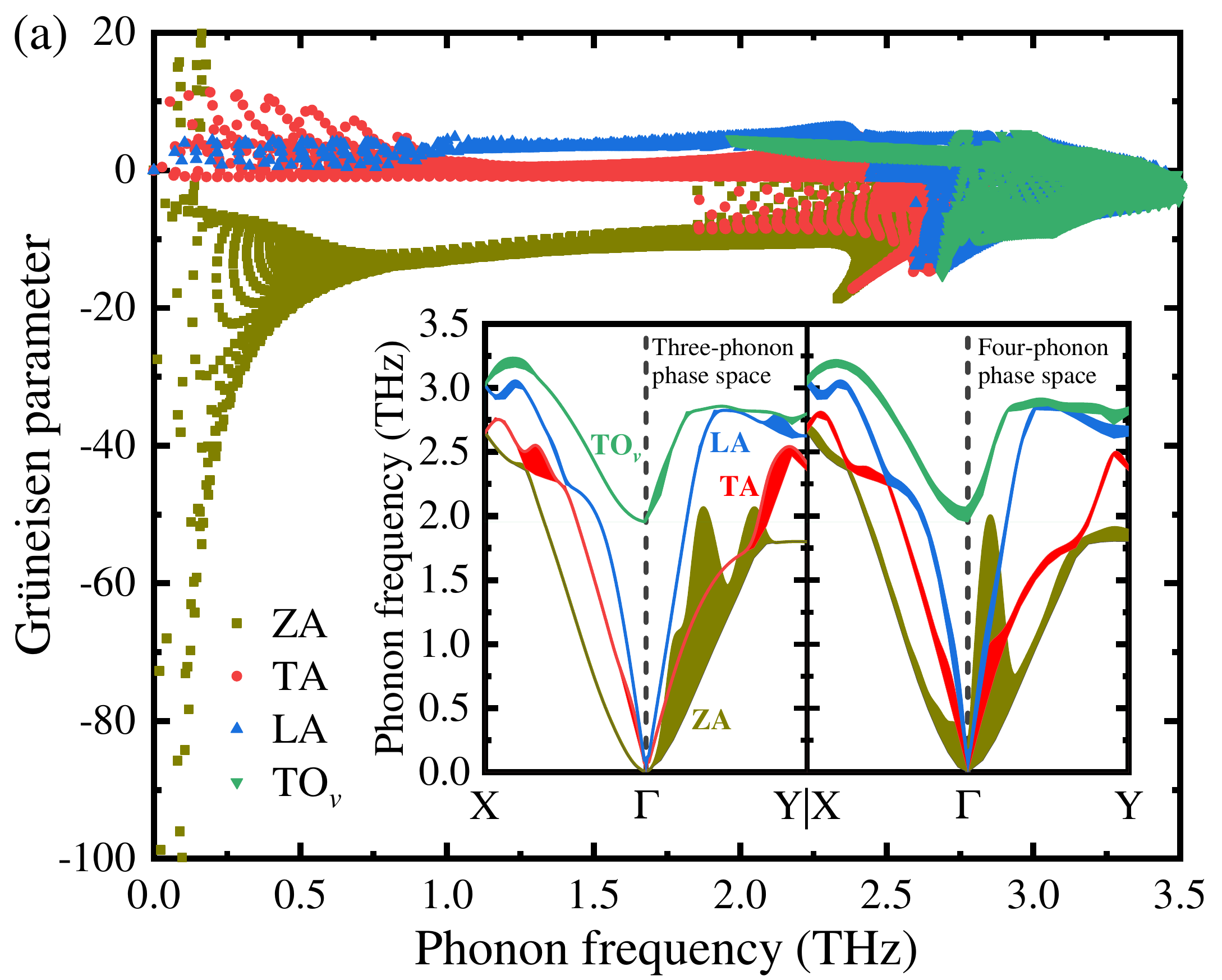}}
\subfigure{\includegraphics[width=7 cm]{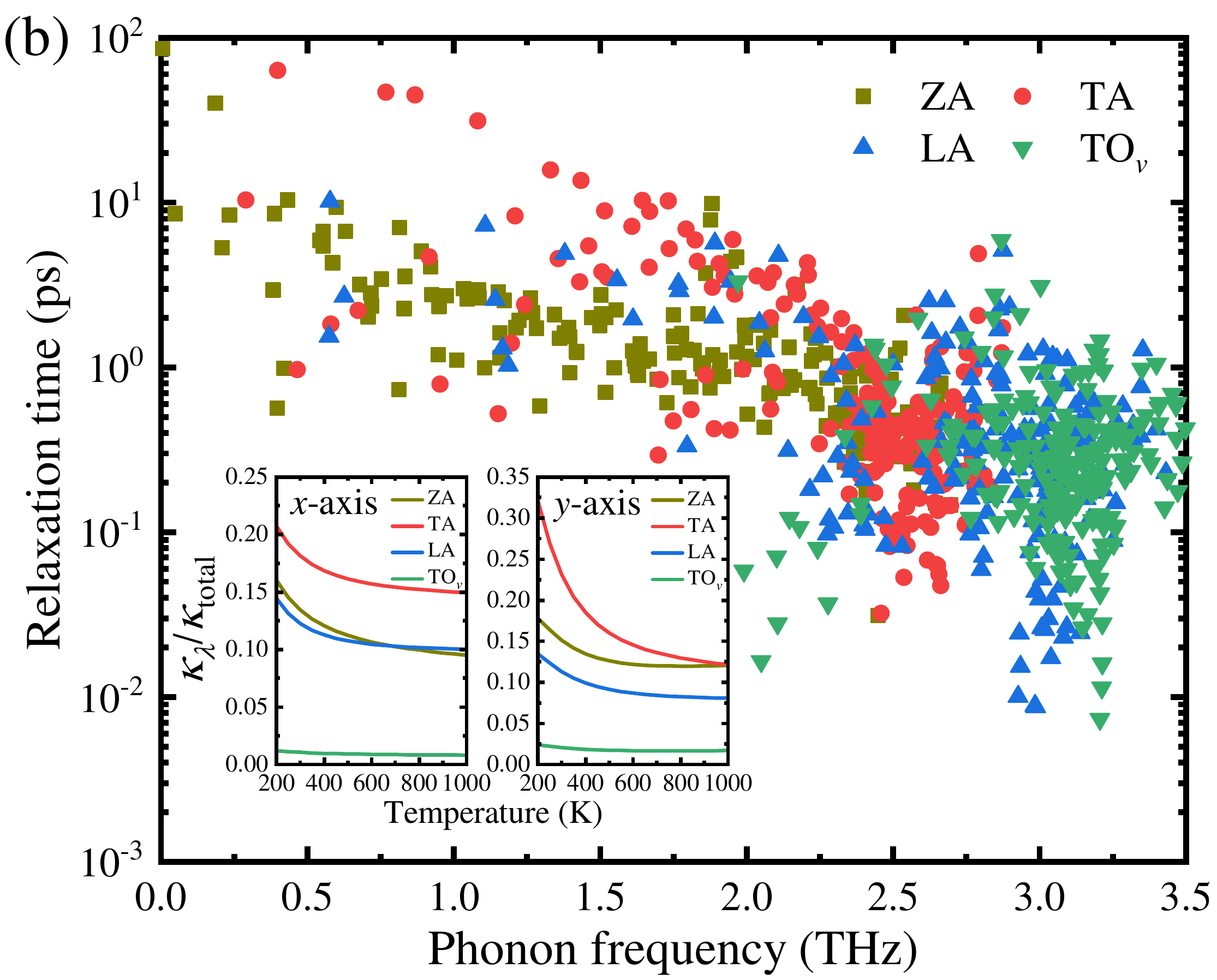}}
\subfigure{\includegraphics[width=7 cm]{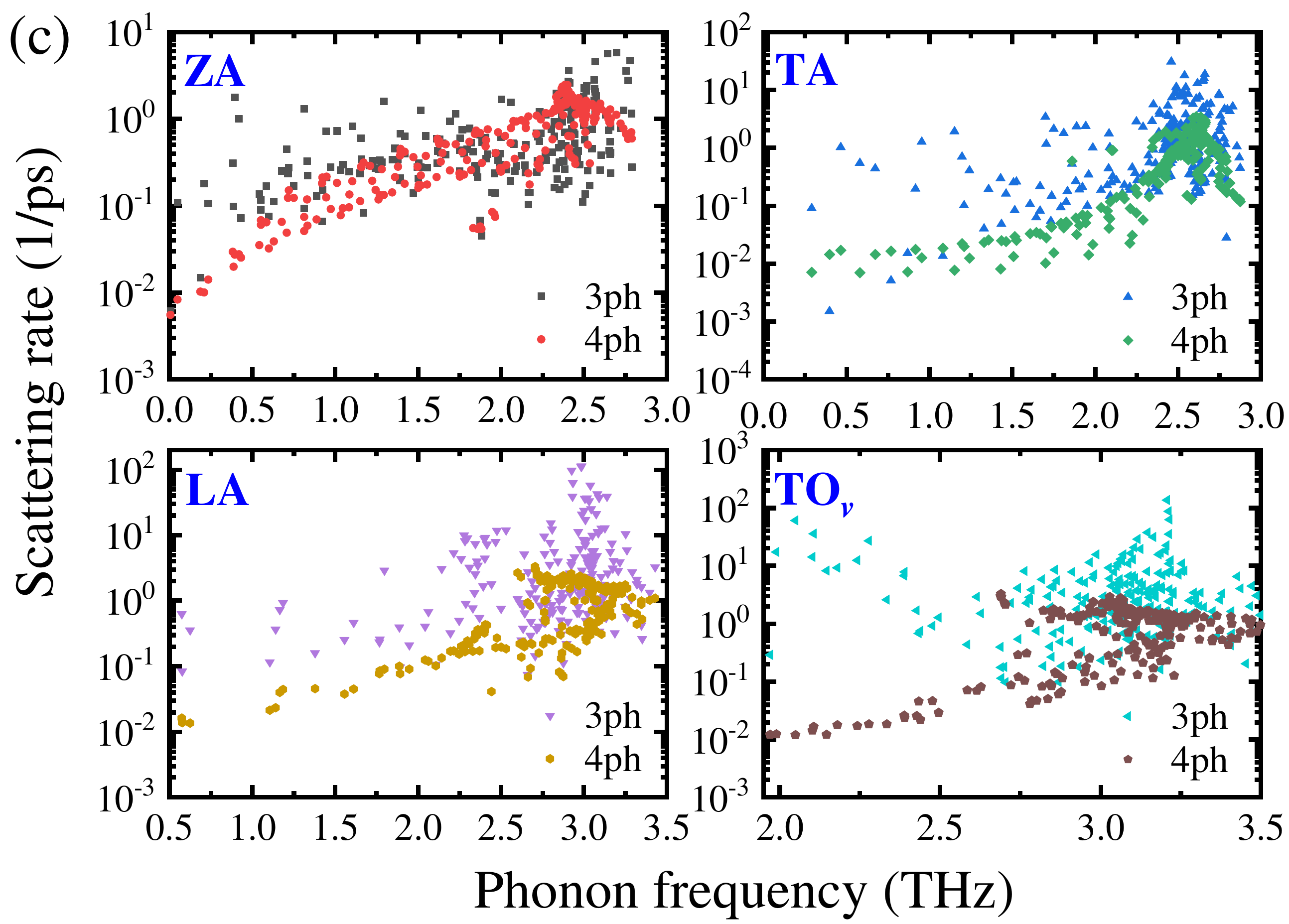}}
\caption{\label{fig:3}
Analysis of the ZA, TA, LA, and TO$_v$ modes: (a) Gr\"{u}neisen paramenters. Inset is the weighted phonon phase space (both 3- and 4-phonon processes) at $T=300$ K. Each branch is shown with a variable-width band equal to the  weighted phonon phase space  in ps$^4$/rad$^4$ \cite{li_shengbte:_2014, li_ultralow_2015}. (b) Relaxation times. Inset shows the normalized contributions to $\kappa_p$. (c) 3- and 4-phonon scattering rates at $T=300$ K.}
\end{figure}

Ultimately, $\kappa_p$ depends both (i) on the anharmonic interaction matrix elements and (ii) on the inverse of phonon phase space volume \cite{li_shengbte:_2014,li_ultralow_2015}.  Regarding (i),  a useful measure of anharmonicity is given by the Gr\"{u}neisen parameter $\gamma$ \cite{li_shengbte:_2014}, which quantifies the volume change with $T$.  Figure \ref{fig:3}(a) shows the calculated $\gamma$ for all acoustic  and TO$_v$ modes of monolayer MoO$_3$. As large $\gamma$ imply large anharmonicity,  we uncover a giant ZA mode anharmonicity. Referring to Fig. \ref{fig:1}(a) inset, ZA is mapped on the deformation of the z-oriented Mo-O bonds which define the mono-layer thickness. It comprises the surface O (in orange in Fig. \ref{fig:1}(a)), which are singly coordinated by a weak double bond with  Mo.  The large $\gamma$ of the ZA enhances the scattering rates, thus contributing to the ultralow $\kappa_p$.  (ii) The phonon phase space  gives a measure of all available phonon scattering processes able to simultaneously satisfy the energy and momentum conservation requirements. The inset of Fig. \ref{fig:3}(a) displays the calculated weighted phonon phase space \cite{li_shengbte:_2014,li_ultralow_2015} of ZA, TA, LA, and TO$_v$ modes. This result confirms  that scattering channels for ZA are abundant.   The phonon phase spaces along $y$ are larger than along $x$, indicating a larger probability for scattering along $y$ and providing a verification for the lower $\kappa_p$ along  $y$,  Fig. \ref{fig:1}(a).  

The decoupled ZA modes of graphene have larger relaxation times than TA and LA ones \cite{lindsay_phonon_2014,liu_continuum_2016}, and  contribute  nearly 75\% to $\kappa_p$. In MoO$_3$, Fig. \ref{fig:3}(b) we find instead  relaxation times of  ZA comparable to those of TA and LA, suggesting instead a strong in-plane and out-of-plane mode coupling. As such, the ZA contribution to $\kappa_{p}$,  see inset of Fig. \ref{fig:3}(b), is of only 13.4\% and 15.1\% along the $x$ and $y$ directions at $T=300$ K, respectively.  

Turning back to Fig. \ref{fig:1}(a), it is important to note the  large difference between $\kappa^{\text{3ph}}_p$ (only the 3-phonon scatterings), and  $\kappa^{\text{3+4ph}}_p$ (with both the 3- and 4-phonon scatterings).  At $T$=300 K, $\kappa^{\text{3ph}}_p$  along the $x$ and $y$ directions are 3.37 and 2.19 W/mK, respectively. These values are further lowered to 1.57 and 1.26 W/mK for  $\kappa^{\text{3+4ph}}_p$, i.e., by 53\% and 42\% along the $x$- and $y$-axis, respectively. The difference grows as $T$ increases,  demonstrating the importance of the 4-phonon scattering in this material. For more evidence on the role of the 4-phonon scatterings, in Fig. \ref{fig:3}(c),  we present the 4-phonon scattering rates of ZA, TA, LA and TO$_v$ modes at $T$=300 K.  These rates are indeed comparable with their 3-phonon scattering ones, which are plotted for a comparison. Only the 4-phonon scattering rate of TO$_v$ is much smaller than its 3-phonon counterpart. Due the failure of energy conservation, TO$_v$ doesn't participate in the 4-phonon processes.


\begin{figure}[htp]
\centering     
\subfigure{\includegraphics[width=7 cm]{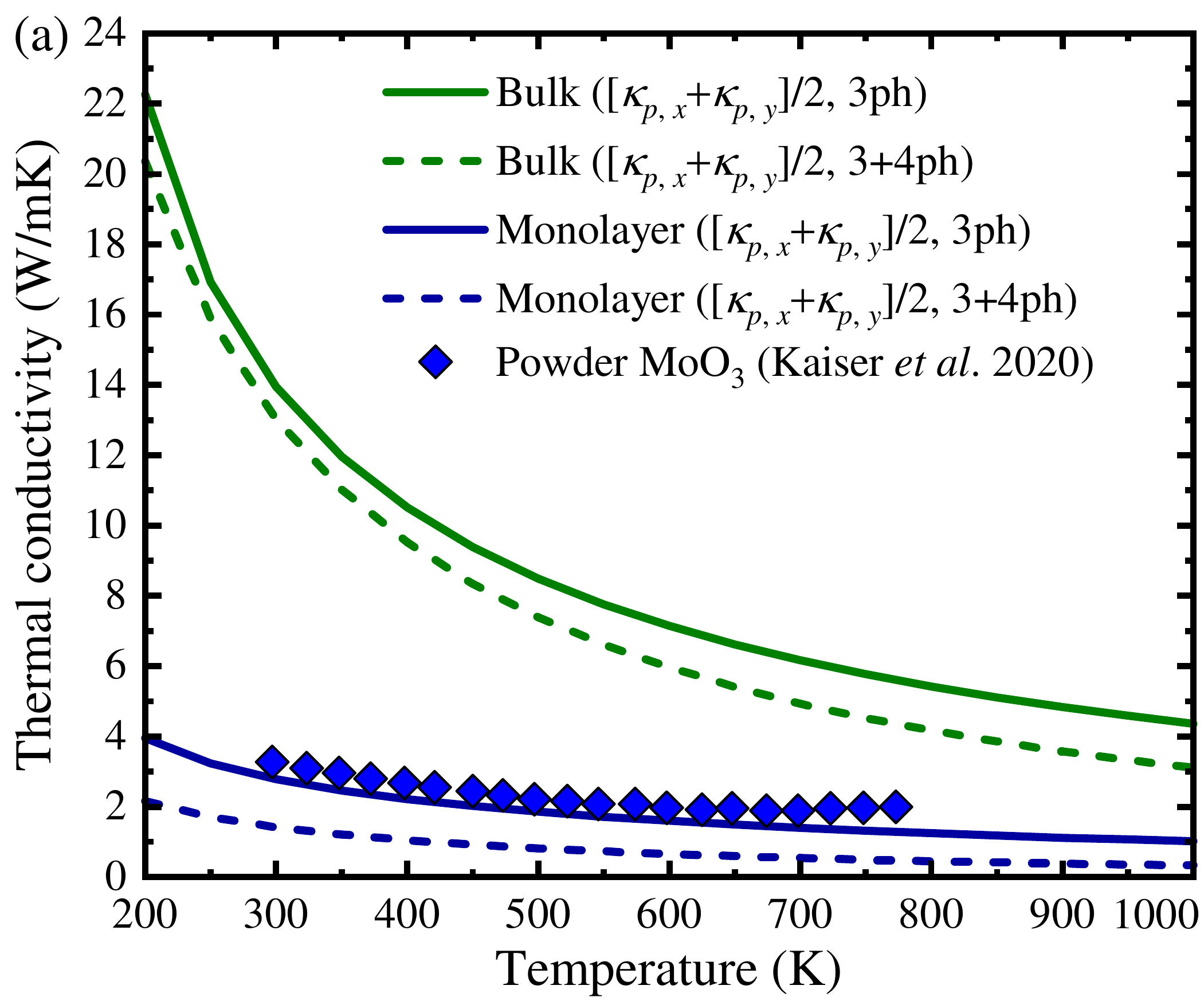}}
\subfigure{\includegraphics[width=7 cm]{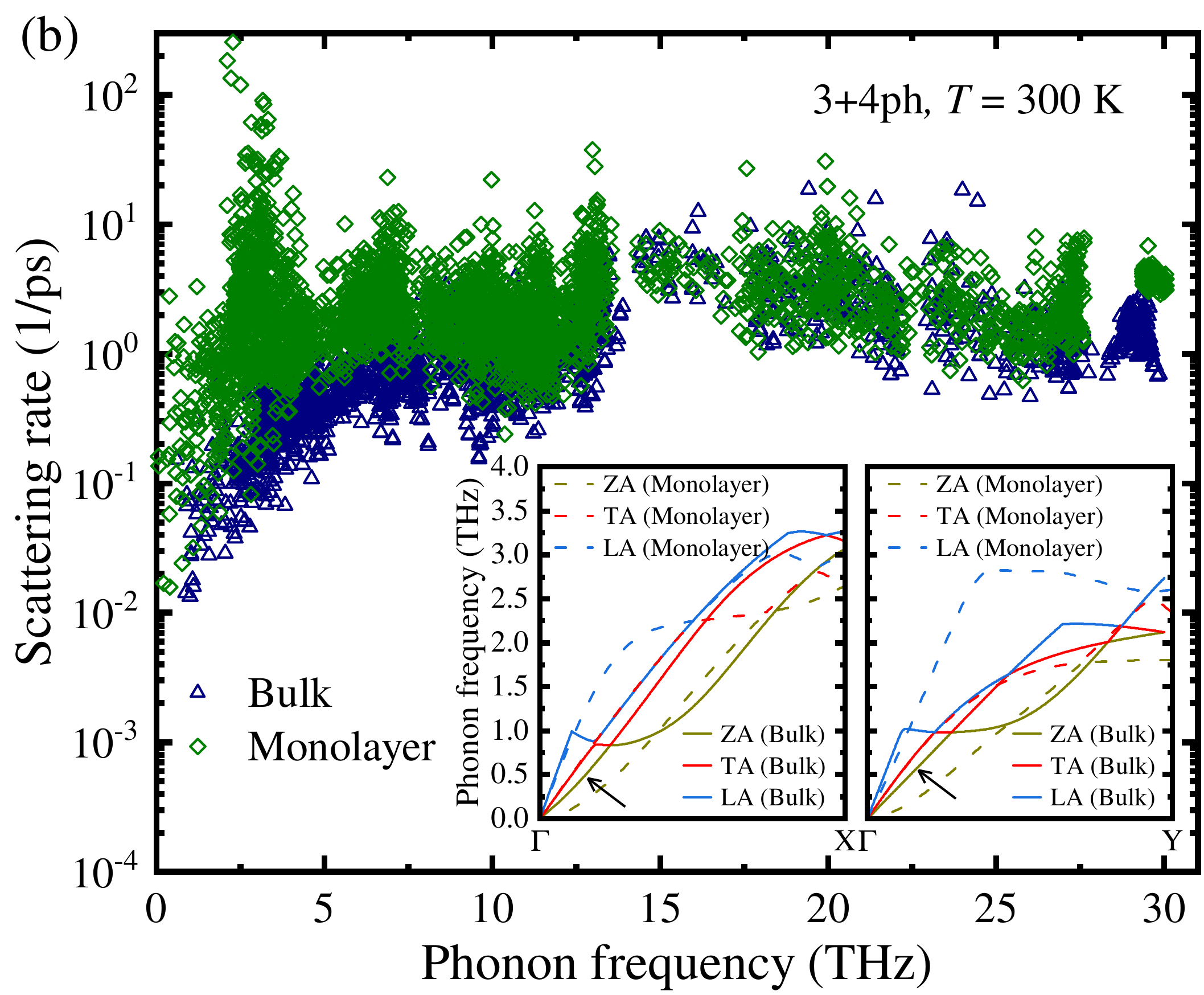}}
\caption{\label{fig:4}
(a) Averaged  $\kappa_p$ of bulk and monolayer MoO$_3$ vs.  $T$.  Both $\kappa_{p}^{\text{3ph}}$ and $\kappa_{p}^{\text{3+4ph}}$ are plotted. Experimental data \cite{kaiser_molybdenum_2020} of powder MoO$_3$ is provided for a comparison. (b) Phonon scattering rates (both 3- and 4-phonon scatterings included)  at  $T=300$ K. Inset shows dispersion of the acoustic modes.}
\end{figure}

For a broader view, Fig. \ref{fig:4}(a) plots our calculated  $\kappa_p$ in MoO$_3$ crystals. The bulk form also exhibits strong anharmonicity which manifests itself in the low values of the averaged in-plane $\kappa_p$ and also in the important contribution of the 4-phonon scatterings.   The larger  $\kappa_p$  of MoO$_3$ crystals in  comparison with the only recently available data \cite{kaiser_molybdenum_2020} is expected as  phonon scatterings on impurity, defects, and grain boundaries of the powder form reduces thermal transport.  However, the larger $\kappa_p$ value of bulk compared to the monolayer is unusual. 2D materials, including graphene \cite{lindsay_flexural_2011} and MoS$_2$ \cite{gu_layer_2016}, were found to have larger $\kappa_p$ than their 3D counterparts. This abnormal behavior can be understood by comparing the phononic scattering rates and dispersions in monolayer and in bulk, Fig. \ref{fig:4}(b). First, the scattering rates in monolayer MoO$_3$ are overall larger than in bulk. Second, there is a difference in the behavior of the ZA modes: In the ultra-flexible monolayer, ZA  are coupled with in-plane acoustic modes to enhance the phonon scattering, which in turn acts to lower $\kappa_p$. In bulk,  the transversal layer vibrations are nevertheless stiffened by the inter-layer interactions:  In the inset of Fig. \ref{fig:4}(b), there is a quartic-to-linear transition (see arrows) as ZA is approaching the $\Gamma$, wheres in monolayer the same mode stays quadratic.  Thus, the flexural modes of the layers have larger group velocity in bulk,  leading to larger $\kappa_p$.  
          

\begin{figure}[htp]
\centerline{\includegraphics[width=7 cm]{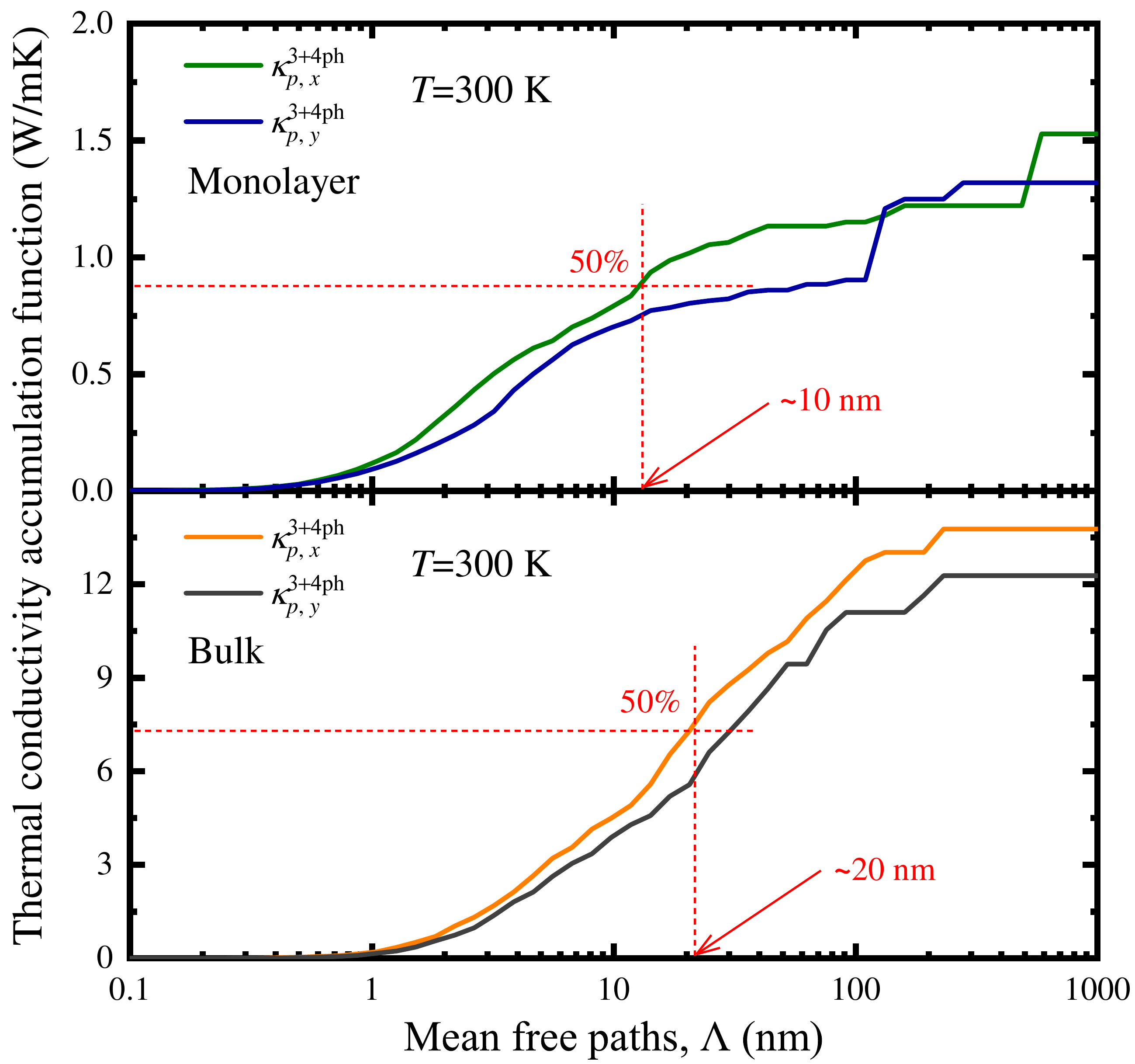}}
\caption{\label{fig:5}
Accumulation functions of $\kappa_p$ along $x$ and $y$ in bulk and monolayer MoO$_3$ vs.  $\Lambda$ at  $T=300$ K.}
\end{figure}

Our predictions can be validated with micro-devices \cite{cai_thermal_2010,gabourie_reduced_2020} for measuring $\kappa_p$, and can motivate MoO$_3$ as a thermoelectric material. As small $D$ is also associated to ultrastrong adhesion \cite{bending1_2011,koenig_ultrastrong_2011}, 2D MoO$_3$ appears suitable for thermal protective layers well adhered to functional nanostructures.  From a practical perspective,  an important consideration is that size reduction to dimensions comparable to the phonon mean free path ($\Lambda$) impacts $\kappa_p$.  The computed the $\kappa_p$ accumulation function, Fig. \ref{fig:5}, which describes the phonon contributions  with  increased $\Lambda$,   indicates that phonons with $\Lambda$  between 1$-$100 nm bring  the dominant contributions. Bulk is more affected by size than the monolayer: In bulk (2D), phonons with $\Lambda <$20 nm (10 nm) are needed to achieve 50\% of $\kappa_p$. 


In conclusion,  monolayer MoO$_3$ is a contender for the 2D material with the lowest $\kappa_p$. The ultralow $\kappa_p$, 1.57 (along $x$) and 1.26 W/mK (along $y$) at $T$=300 K, is attributed to the ideal combination of  small flexural rigidity and soft acoustic modes, which are coupled through the finite layer thickness, and strong intrinsic anharmonicity, which gives large 4-phonon scatterings. Furthermore, $\kappa_p$ of 2D MoO$_3$ is lower than in bulk, an unusual behavior due to a larger phonon scattering rate and lower group velocity of ZA modes in monolayer. Currently, low $\kappa_p$  is tied to buckling of one-atom-thick monolayers \cite{xie_thermal_2014,gao_unusually_2018},  weak interatomic bondings \cite{jain_strongly_2015}, and strong anharmonicity \cite{qin_diverse_2016}.  The mechanism identified  in 2D MoO$_3$ suggests that ultralow $\kappa_p$ can generally be a physical property of (i) few-atoms thick monolayers with  (ii) small $D$, (iii) soft acoustic modes, and (iv) large bonding anharmonicity.

\begin{center}
\textbf{METHODS}
\end{center}

Our calculations employ the projector-augmented-wave (PAW) \cite{kresse_ultrasoft_1999} method based on DFT and density functional perturbation theory (DFPT), as implemented in VASP \cite{kresse_ab_1993}. Due to the failure of the traditional exchange-correlation functionals  in describing vdW, we adopted the optB88 exchange functional \cite{klimes_chemical_2010}. The harmonic and cubic interatomic force constants (2nd- and 3rd-IFCs) for the 3-phonon scattering  calculations were based on  DFTP and finite differencing of DFT forces. The 3rd-IFCs were implemented in thirdorder.py in ShengBTE \cite{li_shengbte:_2014}, which was also used  for the 3-phonon scattering rates. An in-house code \cite{feng_quantum_2016,tong_first-principles_2020} was used for the 4th-IFCs and 4-phonon scattering rate calculations.
\begin{center}
\textbf{SUPPLEMENTARY MATERIAL}
\end{center}
Details on derivation of 3- and 4-phonon terms,  force constants,  structural information of the monolayer, flexural rigidity, and phononic band structure of bulk MoO$_3$.  

\begin{center}
\textbf{ACKNOWLEDGMENTS}
\end{center}

We would like to thank Dr. ChiYung Yam and Dr. Alessandro Pecchia for valuable discussions. Z.T. acknowledges the support by China Postdoctoral Science Foundation (Grant No. 2020M680127), Guangdong Basic and Applied Basic Research Foundation (Grant No. 2020A1515110838), and Shenzhen Science and Technology Program (Grant No. RCBS20200714114919142). T.D. and T.F. acknowledge Mercator Fellowship support from DFG FR-2833/71, and T.F. acknowledges the financial support from the National Natural Science Foundation of China (Grant No. U1930402) as well. Simulations were preformed at the Tainhe2-JK of Beijing Computational Science Research Center (CSRC).

\bibliographystyle{apsrev4_updates}
\bibliography{Reference}


\end{document}